\documentclass[pdflatex,sn-mathphys-num]{sn-jnl}

\usepackage{graphicx}%
\usepackage{multirow}%
\usepackage{amsmath,amssymb,amsfonts}%
\usepackage{amsthm}%
\usepackage{mathrsfs}%
\usepackage[title]{appendix}%
\usepackage{xcolor}%
\usepackage{textcomp}%
\usepackage{manyfoot}%
\usepackage{booktabs}%
\usepackage{algorithm}%
\usepackage{algorithmicx}%
\usepackage{algpseudocode}%
\usepackage{listings}%

\begin{document}

\title[Exciton-polariton stimulated scattering...]{Exciton-polariton stimulated scattering in hybrid halide perovskites}

\author[1]{\fnm{Igor} \sur{Chestnov}}\email{igor\_chestnov@mail.ru}

\author[2]{\fnm{Mikhail} \sur{Masharin}}

\author[1]{\fnm{Valeriy} \sur{Kondratyev}}

\author[3]{\fnm{Ivan} \sur{Iorsh}}

\author[4]{\fnm{Anton} \sur{Samusev}}

\author[1]{\fnm{Anatoly} \sur{Pushkarev}}

\author[1,5]{\fnm{Sergey} \sur{Makarov}}

\author[1,6]{\fnm{Ivan A.} \sur{Shelykh}}

\author*[1,7]{\fnm{Vanik} \sur{Shahnazaryan}}\email{vanikshahnazaryan@gmail.com}

\affil[1]{\orgdiv{Department of Physics}, \orgname{ITMO University}, \orgaddress{ \city{St. Petersburg}, \postcode{197101}, \country{Russia}}}

\affil[2]{ \orgname{Institute of Bioengineering, École Polytechnique Fédérale de Lausanne (EPFL)}, \orgaddress{ \city{Lausanne}, \postcode{1015}, \country{Switzerland}}}

\affil[3]{\orgdiv{Department of Physics, Engineering Physics and Astronomy}, \orgname{Queen’s University}, \orgaddress{ \city{ Kingston, Ontario }, \postcode{K7L 3N6}, \country{Canada}}}

\affil[4]{\orgdiv{Experimentelle Physik 2}, \orgname{Technische Universität Dortmund}, \orgaddress{ \city{Dortmund}, \postcode{44227}, \country{Germany}}}

\affil[5]{\orgdiv{Qingdao Innovation and Development Center}, \orgname{Harbin Engineering University}, \orgaddress{ \city{Qindao}, \postcode{266000}, \country{China}}}

\affil[6]{\orgdiv{Science Institute}, \orgname{University of Iceland}, \orgaddress{ \city{Reykjavik}, \postcode{IS-107}, \country{Iceland}}}

\affil[7]{\orgname{Abrikosov Center for Theoretical Physics}, \orgaddress{ \city{Dolgoprudnyi}, \postcode{141701}, \country{Russia}}}

\abstract{
Halide perovskites, such as methylammonium lead bromide (MAPbBr$_3$), host tightly bound three-dimensional excitons which are robust at room temperature.
Excellent optical properties of MAPbBr$_3$ allow for designing of optical single-mode waveguides and cavities in the frequency range close to the excitonic transitions.
Taken together, this turns MAPbBr$_3$ into an excellent platform for probing exciton-polariton nonlinear phenomena at room temperature. 
Here we investigate ultrafast non-equilibrium dynamics of polaritons under pulsed fs non-resonant excitation.
We demonstrate the presence of the stimulated acoustic phonon-assisted scattering regime above threshold pump fluence, characterized by the explosive growth of emission intensity, a redshift of the emission spectral maximum, spectral narrowing, and sub-picosecond emission dynamics. 
Our theoretical findings are well confirmed by the results of experimental measurements.
}


\maketitle


Exciton-polaritons are half-light half-matter bosonic quasiparticles possessing outstanding many-body properties thanks to their excitonic counterpart
\cite{kavokin2017microcavities}.
The most prominent manifestations of collective behavior of exciton-polaritons is the Bose-Einstein condensation (BEC) \cite{kasprzak2006bose}, and the onset of superfluidity \cite{amo2009superfluidity,amo2011polariton}.
The common scheme of polariton BEC preparation implies a non-resonant excitation, resulting in a formation of high-energy and large momentum dark exciton states, typically referred to as reservoir, with the subsequent transfer to macroscopically occupied coherent ground state \cite{Wouters2007}.  
Such relaxation process is due to exciton scattering on acoustic phonons \cite{piermarocchi1996nonequilibrium}, and the exciton-exciton scattering \cite{tassone1999exciton}.
The efficiency of polariton relaxation is governed by the bosonic nature of excitons, implying the stimulation of scatterings towards occupied states.
The kinetics of exciton-polariton stimulated scattering on acoustic phonons was studied in detail at cryogenic temperatures for conventional semiconducting platforms such as GaAs both theoretically \cite{golub1996energy,tassone1997bottleneck} and experimentally \cite{skolnick2002polariton}.  

Recently lead halide perovskites have attracted a growing interest as an excellent platform for room temperature polaritonics \cite{su2021perovskite}.
Formation of BEC, polariton lasing, and spectrum blueshift was demonstrated in two-dimensional (2D) and quasi one-dimensional structures based on layered and bulk perovskites \cite{su2017room,su2018room,fieramosca2019two,su2020observation,feng2021all,belogur2022theory}.
A different class of hybrid perovskites are methylammonium (MA) lead trihalides such as MAPbI$_3$ and MAPbBr$_3$, which host three-dimensional (3D) excitons with large oscillator strength \cite{soufiani2015polaronic}. 
Despite the large static dielectric constant, here the crystalline softness leads to pronounced polaronic effects, enhancing the exciton binding energy \cite{baranowski2020excitons}.
This results in a stable exciton peak at room temperature in MAPbBr$_3$ \cite{shi2020exciton}.
Here the strong coupling regime was demonstrated in a nanowire geometry \cite{shang2018surface}, in the presence of Bragg reflectors \cite{bouteyre2019room}, and in a photonic crystal slab made of MAPbBr$_3$ \cite{masharin2023room}, and MAPbI$_3$ \cite{masharin2022polaron}.
The study of polaritonic effects in MAPbBr$_3$ includes 
lasing \cite{masharin2023lasing},
ultrafast all-optical modulation \cite{masharin2024giant},
and nonlinear response \cite{masharin2023room}.

The coherent properties of perovskite-based polaritons are magnified in the waveguide geometry supporting long-range polariton propagation where the confinement in the transverse direction is provided by the total internal reflection of the guided optical mode. Hybridization with the waveguide mode results in a steep polariton dispersion that strikingly affects polariton relaxation dynamics which requires an accurate matching of the energy and momentum of the scattering particles.
While the processes of polariton relaxation  \cite{laitz2023uncovering} and parametric scattering \cite{wu2021nonlinear} in hybrid layered perovskites were studied previously, the respective studies for MAPbBr$_3$ are lacking so far. 
Here we investigate the kinetics of exciton-polariton stimulated scatterings in MAPbBr$_3$ slab waveguide.

\begin{figure}[t]
    \centering
    \includegraphics[width = 0.99\linewidth]{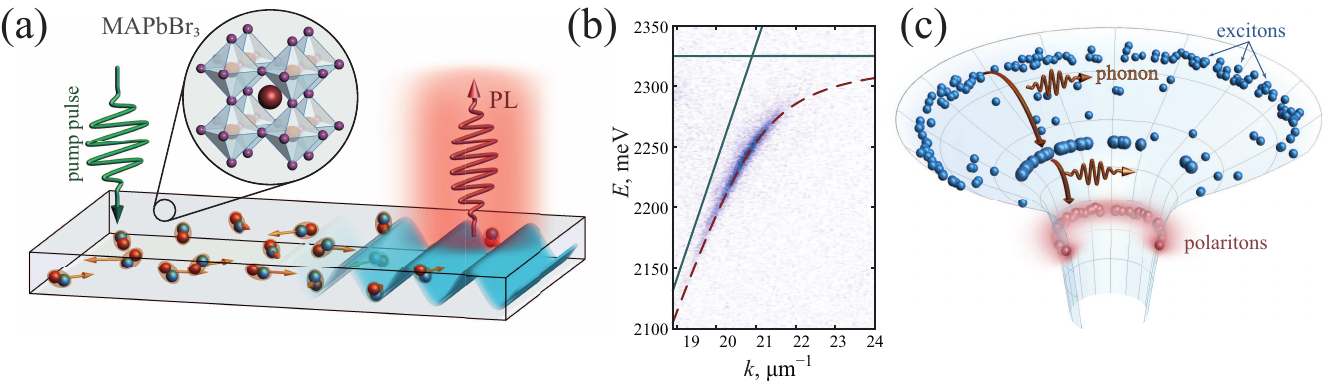}
    \caption{(a) Sketch of the MAPbBr$_3$ slab waveguide. 
    The waveguide mode (blue striped wave) hybridizes with excitons (red and blue balls, corresponding to electron-hole pair) in the strong light-matter coupling regime, forming exciton polaritons.
    (b) Experimentally measured PL spectrum of the sample. The solid lines correspond to exciton and waveguide photon dispersions, and the red dashed curve is the exciton-polariton dispersion, obtained by fitting the PL data via coupled oscillator model. 
    (c) A scheme of phonon-assisted stimulated scattering of polaritons. 
    A non-resonant optical excitation forms a reservoir of large-momentum high-energy dark excitons (blue balls on the top of the dispersion funnel), which rapidly thermalize via scattering on acoustic phonons (red wavy arrows), and form a Boltzmann-like distribution (blue balls in the middle of funnel).
    In the next step the particles scatter to optically active polariton states (red blurry balls), and recombine radiatively.
    }
    \label{fig:sketch}
\end{figure}
\begin{figure}[t]
    \centering
    \includegraphics[width = 0.99\linewidth]{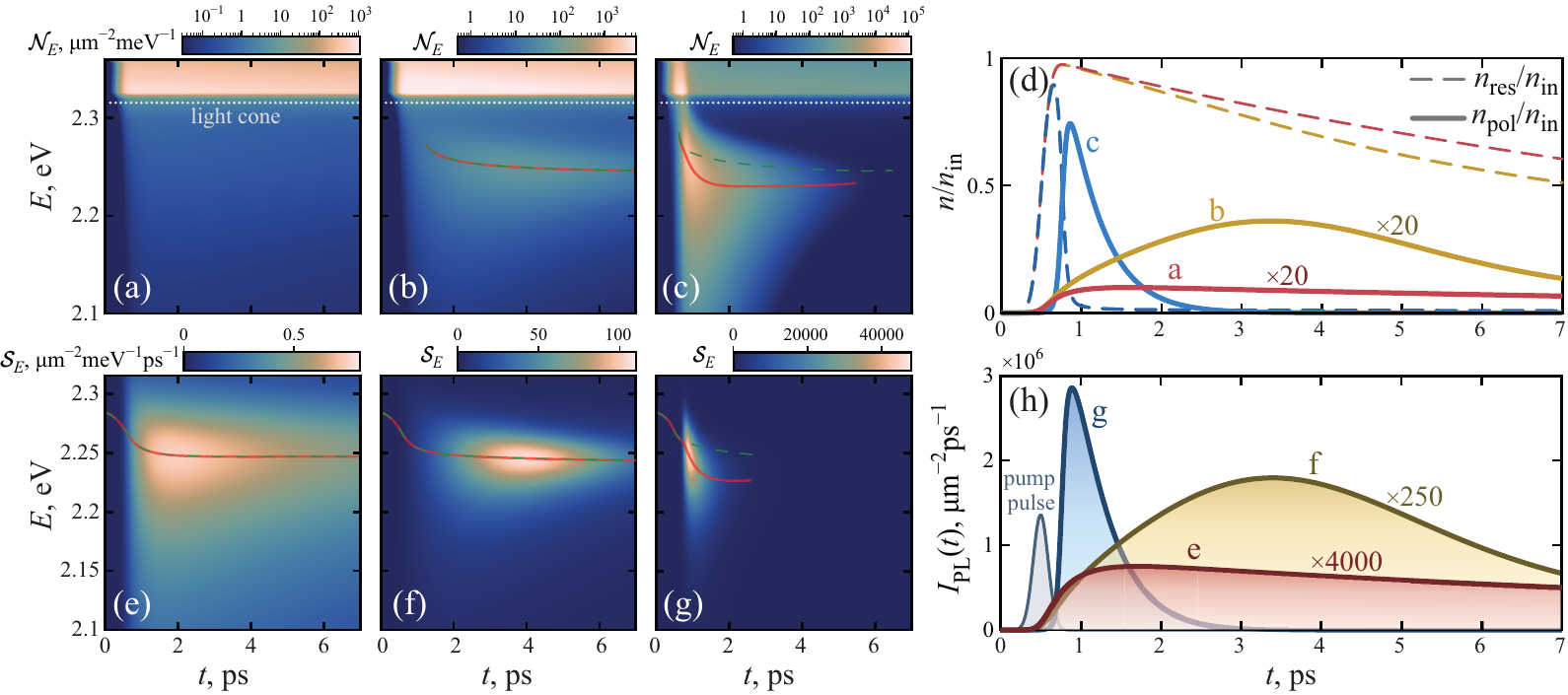}
    \caption{(a-c) Time-resolved evolution of  exciton-polariton population $\mathcal{N}_E$ at elevating pump strength corresponding to the injected exciton densities 
    (a) $n_{\rm in}=2\times10^4$~$\mu$m$^{-2}$, 
    (b) $n_{\rm in}=2\times10^5$~$\mu$m$^{-2}$,  
    (c) $n_{\rm in}=2\times10^6$~$\mu$m$^{-2}$.
    The false-color maps encode an instantaneous particle density $\mathcal{N}_E$ per unit energy range inside the perovskite slab. The white dots separate the energy range of optically active polariton states located within the light cone.
    Increase of the pump power launches the stimulated scattering towards highly-populated states, with subsequent rapid emission, resulting in the formation of population peak in the visible energy domain, shown in (b). 
    In the strongly nonlinear regime (c), a scattering from the peak to the neighboring energy states is enhanced due to their large population, resulting in the gradual redshift of the population maximum.
    Evolution of the spectral maximum is shown with the pale-red lines in (a-c). 
    The same but with the scattering below the light cone turned off is illustrated with the green dashed lines. 
    The radiative losses of the waveguide mode are $\hbar\gamma_{\rm ph}=2.5$~meV.
    (d) Evolution of the energy-integrated population above ($n_{\rm res}$ -- dashed)  and within ($n_{\rm pol}$ -- solid) the light cone normalized with respect to the total injected particle density $n_{\rm in}$.  
    (e-g) Time-resolved dynamics of PL spectrum $\mathcal{S}(E,t)$ at the same values of the pump power as in (a-c). 
    (h) Evolution of the energy-integrated PL $I_{\rm PL}(t)$ at the values of  pump power shown in (e-g).  The arrival time of the above-gap pump pulse of 220~fs FWHM is shown in grey. 
    }
    \label{fig:dynamics}
\end{figure}
\begin{figure}[t]
    \centering
    \includegraphics[width = 0.65\linewidth]{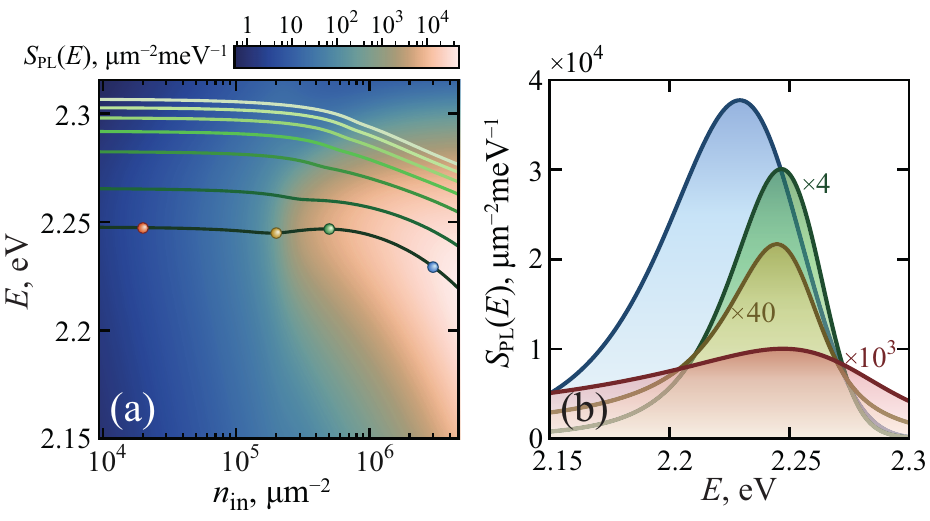}
    \caption{(a) Pump-power dependence of the time-integrated PL spectrum ${S}_{\rm PL}(E)$. 
    The false-color map corresponds to the isolated perovskite waveguide with intrinsic radiative decay $\hbar \gamma_{\rm ph} = 2.5$~meV. 
    The rich-green (lower) line follows position of the spectral maximum. 
    The brighter lines correspond to the PL maximum in the presence of an enhanced photon leakage due to the presence of solid immersion lens used for optical detection (from bottom to the top): 5~meV, 10~meV, 15~meV, 20~meV, 25~meV and 30~meV.
    (b) Time-integrated PL spectrum ${S}_{\rm PL}(E)$ at $\hbar \gamma_{\rm ph} = 2.5$~meV and different pumping strengths shown with colour dots on (a): 
    $n_{\rm in}=3\times 10^6$~$\mu$m$^{-2}$ (blue), 
    $n_{\rm in}=5\times 10^5$~$\mu$m$^{-2}$ (green), $n_{\rm in}=2\times10^5$~$\mu$m$^{-2}$ (yellow) and $n_{\rm in}=2\times 10^4$~$\mu$m$^{-2}$ (red).
   }
    \label{fig:power}
\end{figure}
\begin{figure}[t]
    \centering
    \includegraphics[width = 0.99\linewidth]{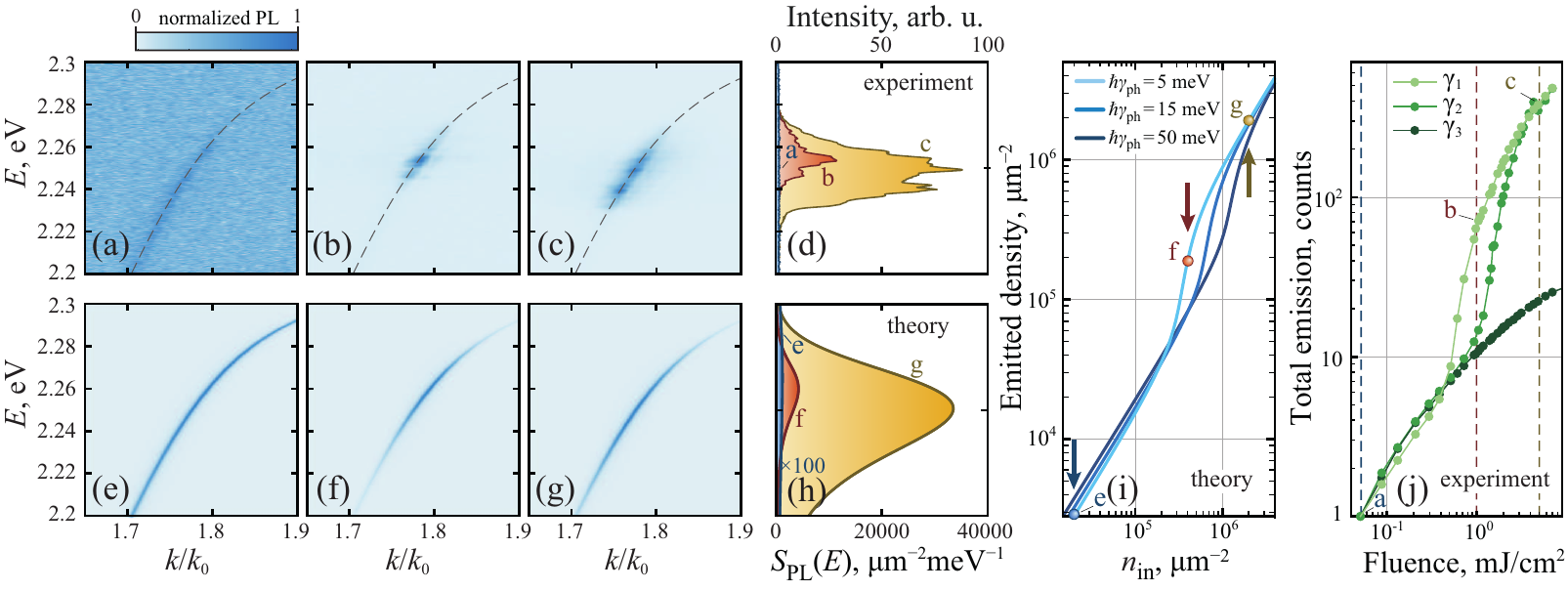}
    \caption{ (a-c) Measured angle-resolved spectra  for the increasing pump fluence: $0.05$~mJ$/{\rm cm}^2$ (a), $1$~mJ$/{\rm cm}^2$ (b) and $5$~mJ$/{\rm cm}^2$ (c). 
    The dashed curve is the fit of the lower polariton branch.
    (d) The measured PL spectra integrated over wavevector for the corresponding pump fluences used in (a-c). 
    (e-g) Theoretical maps of exciton-polariton PL for the varying pump intensity corresponding to
    $n_{\rm in}=2\times10^4$~$\mu$m$^{-2}$~(e), 
    $n_{\rm in}=4\times 10^5$~$\mu$m$^{-2}$~(f) and   
    $n_{\rm in}=2\times10^6$~$\mu$m$^{-2}$~(g) injected exciton densities. The radiative losses of the photon mode are $\hbar\gamma_{\rm ph}=5$~meV.
    (h) Theoretical spectra of PL for pump values used in (e-g).
    (i) The total number of emitted polaritons per unit area $n_{\rm PL}^{\rm tot}$ as a function of the pump power (expressed via the total injected particle density $n_{\rm in}$) at three different values of the photon decay. 
    (j) Integrated experimental emission intensity (total number of detected photons) as a function of pump fluence for three different values of radiative losses (SIL-to-sample distance).}
    \label{fig:experiment}
\end{figure}
%


\section*{Results and Discussion}\label{sec:results}

\textbf{Strong coupling and polariton dispersion}

We study the regime of strong light-matter coupling in a MAPbBr$_3$ thin film, which forms a 2D slab supporting photonic waveguide modes \cite{masharin2024polariton}.
The considered structure is schematically shown in Fig.~\ref{fig:sketch}(a). 
The details of sample fabrication and optical measurements are presented in \hyperref[sec:methods]{Methods} section.
The presence of tightly bound exciton states results in hybridization between the photonic mode and excitons, with the formation of exciton polaritons.
This is confirmed by angle-resolved photoluminescence (PL) measurements shown in Fig.~\ref{fig:sketch} (b), clearly demonstrating an anticrossing feature in the vicinity of exciton resonance and at large momenta lying outside the light cone. 
We fit the PL data to obtain the lower polariton dispersion via coupled oscillator model, $E_{\rm pol} (k)  = \left[ E_{\rm X}(k) + E_{\rm C}(k) -\sqrt{ (2\hbar \Omega)^2 + (E_{\rm X}(k) -E_{\rm C}(k) ) } \right]/2$.
Here $E_{\rm C}(k)$ is the waveguide mode dispersion, 
$E_{\rm X} (k) = E_{\rm X}^{(0)} +\hbar^2 k^2 / (2m_{\rm X}) $ is the exciton dispersion, where $m_{\rm X}$ is the exciton total mass,
$E_{\rm X}^{(0)} = 2325$ meV is the exciton resonance energy \cite{shi2020exciton}, and we set the exciton-photon coupling energy to $\hbar \Omega = 76$ meV.

\vskip 10 pt

\textbf{Polariton nonlinear dynamics and stimulated scattering}

To start with, we address the temporal evolution of the energy-resolved polariton population under the pulsed non-resonant optical excitation, schematically illustrated in Fig.~\ref{fig:sketch}~(c).
The corresponding dynamics is governed by the kinetic equations given by Eq.~\eqref{eq:dynamics} of \hyperref[sec:methods]{Methods} section.
The detailed derivation of the theoretical model is presented in Supplemental Material ({\color{blue}SM}) Section 1.  
The results of numerical simulations are shown in Fig.~\ref{fig:dynamics}.
We define the polariton density per unit energy range per unit area:
\begin{align}
    \label{eq:NE}
    \mathcal{N}_E (t) =   2\pi \rho(E) n_E (t),
\end{align}
which accounts for the polariton density of states $\rho (E) = (2 \partial_{k^2} E)^{-1}$, and $n_E$ is the population of the state with energy $E$.
The pulsed inter-band excitation creates a large density of optically inactive excitons outside of light cone (in what follows referred to as reservoir), as shown in Fig.~\ref{fig:dynamics} (a).
We model this via a pump pulse $P_E(t)$ of Gaussian shape in momentum space, centered around the conduction band edge \cite{tassone1997bottleneck} (see the \hyperref[sec:methods]{Methods}).
Due to scattering on acoustic phonons, particles rapidly redistribute within the reservoir, forming the Boltzmann-like distribution with the bottom at the bottleneck region, defined by an abrupt reduction of the density of states around the inflection point of polariton dispersion.
The vanishing tail of this distribution creates a small polariton population within the optically active region,
with the maximum located at the light cone edge.

Here the reservoir population gradually reduces in time  due to the relatively slow exciton non-radiative decay with a rate $\hbar \gamma_{\rm X}^{\rm nr} = 0.66$ meV, corresponding to 15 ps lifetime.
Within optically active region polaritons additionally decay radiatively, with the speed determined by photonic fraction (Hopfield coefficient $|C_E|^2$), and decay rate of waveguide photons $\hbar \gamma_{\rm ph}= 2.5$ meV, recorded in PL spectra.
Thus, the long-living character of polariton population in Fig.~\ref{fig:dynamics} (a) is due to continuous feeding from the dark reservoir.

We further increase the pump power to hit the nonlinear regime.
Fig.~\ref{fig:dynamics}~(b) shows the evolution of polariton population at the intermediate pumping strength.
Interestingly, the bright exciton-polariton states start to populate after some delay in time, with the population maximum appearing far below the edge of light cone. 
Both the energy distribution and the maximum position of the bright states are determined by the  energy-dependence of the scattering rate. 
The scattering efficiency, in turn, stems from the combination of the exciton scattering rate and the Hopfield coefficients $X_{E}$, $X_{E'}$, defining exciton fractions at the initial and final scattering states within the lower polariton branch, see Eq. \eqref{eq:gE} in \hyperref[sec:methods]{Methods}, and section S2 in {\color{blue} SM}. 
Noteworthy, during the evolution, the population maximum is gradually shifted to lower energy region and saturated subsequently, as indicated by the pale-red line. 
We attribute this shift to the transient process of  thermalization of the reservoir population.
We visualize this further in animated videos illustrating the evolution of population dynamics, available in {\color{blue} SM}.

Fig.~\ref{fig:dynamics} (c) illustrates the polariton population dynamics in a strongly nonlinear regime. 
Here the major part of the excited particles is rapidly scattered towards the bright states, leading to the depletion of reservoir on a sub-picosecond timescale.
The absence of continuous feeding from reservoir determines a rapid decay of polariton density during the subsequent evolution.
Here the intense redshift of population maxima is due to the polariton scatterings within the light cone, enhanced by bosonic stimulation.
To check this, we mimic the evolution turning off the scatterings within the light cone.
The evolution of population maxima for this case is shown with green dashed line in Fig.~\ref{fig:dynamics} (c).
We perform the respective simulations for the linear and intermediate regimes, shown in Fig.~\ref{fig:dynamics} (b) with green dashed line.
The exact match with the pale-red line denoting the evolution of maximal population for the full dynamics indicates the absence of stimulated scatterings within light cone in the linear and intermediate regimes.
 
In Fig.~\ref{fig:dynamics} (d) we show the overall density of dark and bright states, defined as
\begin{align}
     n_{\rm pol} (t) &= \int_0^{E_{\rm lc}} \mathcal{N}_E (t) {\rm d} E, \\
     n_{\rm res} (t) &= \int_{E_{\rm lc}}^\infty \mathcal{N}_E (t) {\rm d} E,
\end{align}
where $E_{\rm lc} = 2315.4$ meV is the energy of the light cone edge defined at the intersection of the polariton dispersion and the light cone of the perovskite, $E_{\rm pol} (k) = \hbar c k / n$.
Here $c$ is the speed of light, and $n=2.3$ is the refractive index of MAPbBr$_3$ \cite{alias2016optical}.
The colors indicate three regimes shown in Figs.~\ref{fig:dynamics} (a), (b), (c).
For easier comparison we normalize the population density with respect to the total injected particle density, defined as
\begin{align}
    n_{\rm in}  =  \int  2\pi \rho(E) P_E (t) {\rm d} E {\rm d} t .
\end{align}
In the linear and intermediate regimes, only a small fraction of excited particles is ultimately scattered into optically active polariton states.
Remarkably, the maximum of overall polariton population in the intermediate regime is reached at later moment as compared to linear regime.
This is due to gradual accumulation of particles, ultimately provoking the stimulated scattering.
In the highly nonlinear regime the stimulated scattering starts almost immediately, determining thus the polariton population maxima at the beginning of evolution.

We next address the PL spectra, which reads
\begin{align}
    \label{eq:SE}
    \mathcal{S}_E(t)  = \gamma_{\rm ph} |C_E|^2 \mathcal{N}_E (t). 
\end{align}
The temporal evolution of PL spectra $\mathcal{S}_E(t)$ for the three regimes of dynamics corresponding to Fig.~\ref{fig:dynamics} (a), (b), (c) is shown in Fig.~\ref{fig:dynamics} (e), (f), (g), respectively.
Interestingly, the spectral distribution of luminescence strongly differs from the particle distribution in the linear regime, c.f. Figs.~\ref{fig:dynamics} (a), (e). 
This is due to the small fraction of photons at the edge of light cone, determined by Hopfield coefficient $|C_E|^2$ in Eq.~\eqref{eq:SE}.
In the intermediate regime the spectral distribution of PL generally follows the particle distribution.
The strongly nonlinear regime is characterized by rapid emission and the large redshift of emission maxima, which are the consequences of reservoir depletion and the stimulated scatterings of polaritons within the light cone.

The temporal evolution of spectrally integrated overall PL, defined as
\begin{align}
     I_{\rm PL} (t) = \int \mathcal{S}_E (t) {\rm d} E, 
\end{align}
is shown in Fig.~\ref{fig:dynamics} (h). 
Here for illustrative purposes we show the profile of the pump pulse.
The evolution of PL generally follows the evolution of polariton density, shown in Fig.~\ref{fig:dynamics}~(d).
Noteworthy, the increase of pump power by two orders of magnitude enhances the PL of about 4000 times, indicating the nonlinear scaling due to the processes of stimulated scattering.

We proceed with examining the pump power dependence of polariton PL.
In Fig.~\ref{fig:power}~(a) the energy-resolved and time-integrated PL spectrum
\begin{align}
    S_{\rm PL} (E) = \int \mathcal{S}_E (t)  {\rm d} t 
\end{align}
versus the pump power is presented.
The spectrum demonstrates a well pronounced threshold-like character. 
Above threshold, the stimulated scattering towards polariton states becomes efficient, resulting in the explosive growth of luminescence intensity.
The rich-green line indicates the spectral position of PL maxima. 
The strong redshift of PL maxima energy with increasing pump power is due to enhanced polariton scatterings within light cone, as discussed above.
Light green lines denote the position of PL maxima for the increased photon decay rate. 
The overall blueshift here is due to intensification of radiative decay as compared to the down-scattering.
Fig.~\ref{fig:power} (b) presents the PL energy cross-sections for several selected pump powers, corresponding to Fig.~\ref{fig:power} (a).
The enhancement of PL intensity, the redshift of maxima, and the spectral narrowing with increasing pump power are due to the polariton stimulated scattering.


\vskip 10 pt
\textbf{Photoluminescence spectra}

We next address experimentally the polariton nonlinear dynamics via angle-resolved spectroscopy at room temperature with solid immersion lens (SIL), as discussed in \hyperref[sec:methods]{Methods}. 
The presence of SIL in the vicinity of the perovskite waveguide allows measuring the momentum-resolved PL intensity of the waveguide mode below the vacuum light cone through the evanescence tails coupled to the SIL \cite{kondratyev2023probing}. 
In addition, coupling to the SIL enhances the radiative losses of the waveguide mode.
The energy and momentum resolved experimental PL spectra for three values of pulsed pump fluence are shown in Fig.~\ref{fig:experiment}~(a), (b), (c). 
Here the in-plane momentum is scaled to units of $k_0 = E_{\rm X}^{(0)}/(\hbar c)$. 
At low pump fluence the PL spectra of the waveguide mode is shown in Fig.~\ref{fig:experiment} (a).
The increase of pump fluence results in spectral narrowing and strong intensity increase of PL due to the stimulated relaxation, c.f. Fig.~\ref{fig:experiment} (a), (b).
Further increase of pump fluence leads to a redshift of the PL spectral maximum, as shown in Fig.~\ref{fig:experiment} (c).
The experimental PL spectra  for the discussed regimes are summarized in Fig.~\ref{fig:experiment}~(d).

We next model theoretically the PL spectra. 
Here we choose the photon radiative decay $\hbar \gamma_{\rm ph } =5$ meV, taking into account the impact of SIL.
The resulting time-integrated PL spectra $S_{\rm PL} (E)$ with momentum resolution are shown in Figs.~\ref{fig:experiment} (e), (f), (g)  for the input densities 
$n_{\rm in}=2\times10^4$~$\mu$m$^{-2}$,
$n_{\rm in}=4\times 10^5$~$\mu$m$^{-2}$, 
$n_{\rm in}=2\times10^6$~$\mu$m$^{-2}$,
respectively.
Fig.~\ref{fig:experiment} (h) summarizes the calculated PL spectra $S_{\rm PL} (E)$ for the three values of pump intensity.
The results of numerical modeling are in good agreement with experimental data.

The calculated emission intensity can be quantified via overall emitted polariton density, reading as
\begin{align}
     n_{\rm PL}^{\rm tot} = \int I_{\rm PL} (t) {\rm d} t.
\end{align}
The dependence of emitted polariton density $n_{\rm PL}^{\rm tot}$ on the input rate $n_{\rm in}$
for three values of photon decay rate is shown in Fig.~\ref{fig:experiment} (i).
The pronounced threshold-like scaling of emitted density is due to the initiation of the stimulated scattering towards polariton states above threshold pump power. 
The respective experimentally measured emission intensity versus the pump fluence is shown in Fig.~\ref{fig:experiment} (j).
Here three curves correspond to different values of photon radiative decay, defined by the SIL-to-sample distance.
The calculated density of emitted polaritons is in excellent agreement with the measured emission intensity.
Both experimental and theoretical data confirm that the increase of polariton radiative losses leads to increase of the stimulated emission threshold provided by the reduction of the particle densities for corresponding pump fluences.

\section*{Conclusion}\label{sec:conclusion}

In summary, we have investigated the ultra-fast relaxation dynamics of non-resonantly excited self-hybridized exciton-polaritons in a thin film of MAPbBr$_3$. 
Unlike the planar microcavity case, the waveguide-based polaritons lack the ground state.
That is why the nonlinear relaxation enabled by stimulated scattering results in the gradual redshift of the PL spectrum towards the linear part of polariton dispersion. 
The corresponding temporal evolution is manifested in a fast diffusion of the polariton spectral density resulting in the effective broadening of PL spectrum. 
An onset of the stimulated regime of the acoustic phonon-assisted scattering is accompanied by the threshold-like behaviour of the total PL intensity. 
At a critical pump fluence the total PL intensity exhibits a rapid growth. 
The threshold of pump fluence can be controlled via tuning the waveguide mode decay rate.
Our findings pave the way to harnessing exciton relaxation dynamics for engineering ultra-fast light-emitting devices.

\section*{Methods}\label{sec:methods}

\textbf{Model}

We model the time and energy resolved dynamics of polariton population with the following set of kinetic equations:
\begin{align}
    \label{eq:dynamics}
    \frac{{\rm d} n_{ E } }{{\rm d}  t} & =  P_{ E }( t)
    - \gamma_E n_{E} \notag \\
    &+  W_0  \delta  E 
    \sum_{E < E'}  g(E, E') 
    \left[  n_{E'} (n_{E}+1)
    -e^{-\Delta E/(k_B T) }
    n_{E} (n_{E'}+1)
    \right] \notag \\
    &+  W_0 \delta  E 
    \sum_{E > E'}  g(E, E')
    \left[ e^{-\Delta E/(k_B T) } n_{E'} (n_{E}+1)
    - n_{ E } (n_{ E '}+1)
    \right] 
\end{align}
for the population $n_{E}$ of the momentum state with energy $E$.
Derivation of Eq.~\eqref{eq:dynamics} is presented in {\color{blue}SM} (see also Refs. \cite{piermarocchi1996nonequilibrium,golub1996energy,tassone1997bottleneck,tassone1999exciton}).
Here the decay rate is $\gamma_E = |C_E|^2 \gamma_{\rm ph}  + |X_E|^2 \gamma_{\rm X}^{\rm nr}$,
and 
\begin{align}
    X_E = \frac{1}{\sqrt{2}}
    \sqrt{1- \frac{E_{\rm C}(E)-E_{\rm X}}{\sqrt{[E_{\rm C}(E)-E_{\rm X}]^2 + 4 \Omega^2} }  }, 
    \quad C_E =\sqrt{1-|X_E|^2}
\end{align}
are the Hopfield coefficients.
The coefficient 
\begin{align}
    W_0 = \frac{\left|D_{\rm c} - D_{\rm v}\right|^2}{(2\pi)^2 \rho_m s^4 \hbar^3 }  ,
\end{align}
defines the scattering amplitude with $\rho_m = 3808$ kg$/$m$^3$ being the material mass density, and $s= 2987$ m$/$s -- the sound velocity \cite{guo2017polar};
$D_c$, $D_v$ are the deformation constants of the conduction and valence bands, respectively, and
we estimate $|D_c - D_v| \sim 3$ eV \cite{butler2015band,mante2017electron}.
The scattering rate between the initial $E^\prime$ and the target $E$ states reads
\begin{equation}
    \label{eq:gE}
    g(E,E') =  \left|X_{E}\right|^2  \left|X_{E^\prime}\right|^2  \frac{ \Delta E^2  }{\partial_{k'^2} E'} 
    \frac{ F\left( \frac{\theta_{\rm max}}{2}, 
    \frac{4k'_{ E'} k_{E}}
    {(\Delta E/\hbar s)^2 -  (k_{E} - k'_{E'})^2 } \right)  }
    { \sqrt{  (\Delta E/\hbar s)^2 -  (k_{E} - k'_{E'})^2 } } 
    \left( \frac{1}{e^{\Delta E/(k_B T) }-1}  +1 \right),
\end{equation}
where $F$ is the elliptic integral of the first kind, $\Delta E = |E - E'|$, $k_E$, $k'_{E'}$ are the respective wave vectors' amplitude, $k_B$ is the Boltzmann constant, $T=300$ K denotes the 
temperature.

We assume that the nonresonant pump creates excitons with the energy $E = E_{\rm X}^{(0)} + E_{\rm b}$ close to the conduction band edge, where $E_{\rm b} = 35$~meV is the exciton binding energy in MAPbBr$_3$ \cite{soufiani2015polaronic,shi2020exciton,masharin2023room}. 
In particular, we take a Gaussian pulse
\begin{align}
    P = P_0 \exp\left(-\frac{c^2 m_{\rm X} }{ E_{\rm X}^{(0)2} } \left( \sqrt{E_{\rm X} (k) -E_{\rm X}^{(0)}   } -\sqrt{E_{\rm b} } \right)^2 \right) 
    \exp\left(- \frac{(t-t_0)^2}{2\tau^2} \right),
\end{align}
where
$\tau = 93$~fs is the pulse duration corresponding to 220~fs full width at half maximum (FWHM).



\textbf{Sample fabrication}

A MAPbBr$_3$ thin film was fabricated using the spin-coating method \cite{jeon2014solvent}. In a nitrogen dry box, 56.0 mg of methylammonium bromide (MABr) from GreatCell Solar and 183.5 mg of Lead(II) bromide (PbBr$_2$) from TCI were mixed in a glass vial. These salts were dissolved in a 1 mL DMF solvent mixture (3:1 ratio), resulting in a 0.5M MAPbBr$_3$ solution, which was stirred for 24 hours at room temperature.

SiO$_2$ substrates (12.5 × 12.5 mm) were cleaned by sonication in deionized water, acetone, and 2-propanol for 10 minutes each, followed by 10 minutes in an oxygen plasma cleaner. The following spin-coating procedure was performed in the dry nitrogen glovebox. 30 µL of the MAPbBr$_3$ solution was applied to the substrate, which was then spun at 3,000 rpm for 40 seconds. After 25 seconds, 300 µL of toluene was dripped onto the spinning substrate. Finally, the spin-coated substrates were annealed at 90°C for 10 minutes.

\textbf{Optical measurements}

Angle-resolved measurements of the perovskite film's guided mode dispersion below the light line were conducted using a back-focal-plane imaging and spectroscopy setup with a high refractive index ZnSe solid immersion lens (SIL) evanescently coupled to the sample\cite{kondratyev2023probing}. The spectroscopy was carried out using an imaging spectrometer equipped with a liquid-nitrogen-cooled CCD camera (Princeton Instruments SP2500+PyLoN). A Mitutoyo 100x VIS HR objective (NA = 0.9) with a ZnSe SIL granted an effective numerical aperture of $\approx$2.25. The SIL-to-sample distance (air gap) below 100 nm was precisely controlled with a piezo positioner. 
 Bringing the SIL closer to the sample surface allowed us to increase the radiative losses of the studied modes. 
 For pump-dependent measurements of angle-resolved emission spectra, the sample was excited by 515 nm femtosecond pulses with a 100 kHz repetition rate (Pharos, Light Conversion). 

\backmatter

\bmhead{Supplementary information}
Online supplementary information is available at ... .

\bmhead{Acknowledgements}

The research is supported by the Ministry of Science and Higher Education of the Russian
Federation (Goszadaniye) Project No. FSMG-2023-0011.
V.S. acknowledges the support of “Basis” Foundation (Project No. 22-1-3-43-1).



\begin{itemize}
\item Author contribution.
V.S., I.C., I.I., and I.A.S. developed the theoretical model.
I.C. conducted the numerical simulations. 
M.M., A.P., and S.M. fabricated the sample.
M.M., V.K., and A.S. performed the spectroscopic measurements.
V.S. managed the project.
All authors extensively discussed the results and  participated in editing of the manuscript.
\end{itemize}








\begin{thebibliography}{35}
\ifx \bisbn   \undefined \def \bisbn  #1{ISBN #1}\fi
\ifx \binits  \undefined \def \binits#1{#1}\fi
\ifx \bauthor  \undefined \def \bauthor#1{#1}\fi
\ifx \batitle  \undefined \def \batitle#1{#1}\fi
\ifx \bjtitle  \undefined \def \bjtitle#1{#1}\fi
\ifx \bvolume  \undefined \def \bvolume#1{\textbf{#1}}\fi
\ifx \byear  \undefined \def \byear#1{#1}\fi
\ifx \bissue  \undefined \def \bissue#1{#1}\fi
\ifx \bfpage  \undefined \def \bfpage#1{#1}\fi
\ifx \blpage  \undefined \def \blpage #1{#1}\fi
\ifx \burl  \undefined \def \burl#1{\textsf{#1}}\fi
\ifx \doiurl  \undefined \def \doiurl#1{\url{https://doi.org/#1}}\fi
\ifx \betal  \undefined \def \betal{\textit{et al.}}\fi
\ifx \binstitute  \undefined \def \binstitute#1{#1}\fi
\ifx \binstitutionaled  \undefined \def \binstitutionaled#1{#1}\fi
\ifx \bctitle  \undefined \def \bctitle#1{#1}\fi
\ifx \beditor  \undefined \def \beditor#1{#1}\fi
\ifx \bpublisher  \undefined \def \bpublisher#1{#1}\fi
\ifx \bbtitle  \undefined \def \bbtitle#1{#1}\fi
\ifx \bedition  \undefined \def \bedition#1{#1}\fi
\ifx \bseriesno  \undefined \def \bseriesno#1{#1}\fi
\ifx \blocation  \undefined \def \blocation#1{#1}\fi
\ifx \bsertitle  \undefined \def \bsertitle#1{#1}\fi
\ifx \bsnm \undefined \def \bsnm#1{#1}\fi
\ifx \bsuffix \undefined \def \bsuffix#1{#1}\fi
\ifx \bparticle \undefined \def \bparticle#1{#1}\fi
\ifx \barticle \undefined \def \barticle#1{#1}\fi
\bibcommenthead
\ifx \bconfdate \undefined \def \bconfdate #1{#1}\fi
\ifx \botherref \undefined \def \botherref #1{#1}\fi
\ifx \url \undefined \def \url#1{\textsf{#1}}\fi
\ifx \bchapter \undefined \def \bchapter#1{#1}\fi
\ifx \bbook \undefined \def \bbook#1{#1}\fi
\ifx \bcomment \undefined \def \bcomment#1{#1}\fi
\ifx \oauthor \undefined \def \oauthor#1{#1}\fi
\ifx \citeauthoryear \undefined \def \citeauthoryear#1{#1}\fi
\ifx \endbibitem  \undefined \def \endbibitem {}\fi
\ifx \bconflocation  \undefined \def \bconflocation#1{#1}\fi
\ifx \arxivurl  \undefined \def \arxivurl#1{\textsf{#1}}\fi
\csname PreBibitemsHook\endcsname

\bibitem[\protect\citeauthoryear{Kavokin et~al.}{2017}]{kavokin2017microcavities}
\begin{bbook}
\bauthor{\bsnm{Kavokin}, \binits{A.}},
\bauthor{\bsnm{Baumberg}, \binits{J.J.}},
\bauthor{\bsnm{Malpuech}, \binits{G.}},
\bauthor{\bsnm{Laussy}, \binits{F.P.}}:
\bbtitle{Microcavities}.
\bpublisher{Oxford university press}, \blocation{Oxford, UK}
(\byear{2017})
\end{bbook}
\endbibitem

\bibitem[\protect\citeauthoryear{Kasprzak et~al.}{2006}]{kasprzak2006bose}
\begin{barticle}
\bauthor{\bsnm{Kasprzak}, \binits{J.}},
\bauthor{\bsnm{Richard}, \binits{M.}},
\bauthor{\bsnm{Kundermann}, \binits{S.}},
\bauthor{\bsnm{Baas}, \binits{A.}},
\bauthor{\bsnm{Jeambrun}, \binits{P.}},
\bauthor{\bsnm{Keeling}, \binits{J.M.J.}},
\bauthor{\bsnm{Marchetti}, \binits{F.}},
\bauthor{\bsnm{Szyma{\'n}ska}, \binits{M.}},
\bauthor{\bsnm{Andr{\'e}}, \binits{R.}},
\bauthor{\bsnm{Staehli}, \binits{J.}}, \betal:
\batitle{Bose--einstein condensation of exciton polaritons}.
\bjtitle{Nature}
\bvolume{443}(\bissue{7110}),
\bfpage{409}--\blpage{414}
(\byear{2006})
\end{barticle}
\endbibitem

\bibitem[\protect\citeauthoryear{Amo et~al.}{2009}]{amo2009superfluidity}
\begin{barticle}
\bauthor{\bsnm{Amo}, \binits{A.}},
\bauthor{\bsnm{Lefr{\`e}re}, \binits{J.}},
\bauthor{\bsnm{Pigeon}, \binits{S.}},
\bauthor{\bsnm{Adrados}, \binits{C.}},
\bauthor{\bsnm{Ciuti}, \binits{C.}},
\bauthor{\bsnm{Carusotto}, \binits{I.}},
\bauthor{\bsnm{Houdr{\'e}}, \binits{R.}},
\bauthor{\bsnm{Giacobino}, \binits{E.}},
\bauthor{\bsnm{Bramati}, \binits{A.}}:
\batitle{Superfluidity of polaritons in semiconductor microcavities}.
\bjtitle{Nature Physics}
\bvolume{5}(\bissue{11}),
\bfpage{805}--\blpage{810}
(\byear{2009})
\end{barticle}
\endbibitem

\bibitem[\protect\citeauthoryear{Amo et~al.}{2011}]{amo2011polariton}
\begin{barticle}
\bauthor{\bsnm{Amo}, \binits{A.}},
\bauthor{\bsnm{Pigeon}, \binits{S.}},
\bauthor{\bsnm{Sanvitto}, \binits{D.}},
\bauthor{\bsnm{Sala}, \binits{V.}},
\bauthor{\bsnm{Hivet}, \binits{R.}},
\bauthor{\bsnm{Carusotto}, \binits{I.}},
\bauthor{\bsnm{Pisanello}, \binits{F.}},
\bauthor{\bsnm{Lem{\'e}nager}, \binits{G.}},
\bauthor{\bsnm{Houdr{\'e}}, \binits{R.}},
\bauthor{\bsnm{Giacobino}, \binits{E.}}, \betal:
\batitle{Polariton superfluids reveal quantum hydrodynamic solitons}.
\bjtitle{Science}
\bvolume{332}(\bissue{6034}),
\bfpage{1167}--\blpage{1170}
(\byear{2011})
\end{barticle}
\endbibitem

\bibitem[\protect\citeauthoryear{Wouters and Carusotto}{2007}]{Wouters2007}
\begin{barticle}
\bauthor{\bsnm{Wouters}, \binits{M.}},
\bauthor{\bsnm{Carusotto}, \binits{I.}}:
\batitle{Excitations in a nonequilibrium bose-einstein condensate of exciton polaritons}.
\bjtitle{Phys. Rev. Lett.}
\bvolume{99},
\bfpage{140402}
(\byear{2007})
\end{barticle}
\endbibitem

\bibitem[\protect\citeauthoryear{Piermarocchi et~al.}{1996}]{piermarocchi1996nonequilibrium}
\begin{barticle}
\bauthor{\bsnm{Piermarocchi}, \binits{C.}},
\bauthor{\bsnm{Tassone}, \binits{F.}},
\bauthor{\bsnm{Savona}, \binits{V.}},
\bauthor{\bsnm{Quattropani}, \binits{A.}},
\bauthor{\bsnm{Schwendimann}, \binits{P.}}:
\batitle{Nonequilibrium dynamics of free quantum-well excitons in time-resolved photoluminescence}.
\bjtitle{Physical Review B}
\bvolume{53}(\bissue{23}),
\bfpage{15834}
(\byear{1996})
\end{barticle}
\endbibitem

\bibitem[\protect\citeauthoryear{Tassone and Yamamoto}{1999}]{tassone1999exciton}
\begin{barticle}
\bauthor{\bsnm{Tassone}, \binits{F.}},
\bauthor{\bsnm{Yamamoto}, \binits{Y.}}:
\batitle{Exciton-exciton scattering dynamics in a semiconductor microcavity and stimulated scattering into polaritons}.
\bjtitle{Physical Review B}
\bvolume{59}(\bissue{16}),
\bfpage{10830}
(\byear{1999})
\end{barticle}
\endbibitem

\bibitem[\protect\citeauthoryear{Golub et~al.}{1996}]{golub1996energy}
\begin{barticle}
\bauthor{\bsnm{Golub}, \binits{L.}},
\bauthor{\bsnm{Scherbakov}, \binits{A.}},
\bauthor{\bsnm{Akimov}, \binits{A.}}:
\batitle{Energy distributions of 2d excitons in the presence of nonequilibrium phonons}.
\bjtitle{Journal of Physics: Condensed Matter}
\bvolume{8}(\bissue{13}),
\bfpage{2163}
(\byear{1996})
\end{barticle}
\endbibitem

\bibitem[\protect\citeauthoryear{Tassone et~al.}{1997}]{tassone1997bottleneck}
\begin{barticle}
\bauthor{\bsnm{Tassone}, \binits{F.}},
\bauthor{\bsnm{Piermarocchi}, \binits{C.}},
\bauthor{\bsnm{Savona}, \binits{V.}},
\bauthor{\bsnm{Quattropani}, \binits{A.}},
\bauthor{\bsnm{Schwendimann}, \binits{P.}}:
\batitle{Bottleneck effects in the relaxation and photoluminescence of microcavity polaritons}.
\bjtitle{Physical Review B}
\bvolume{56}(\bissue{12}),
\bfpage{7554}
(\byear{1997})
\end{barticle}
\endbibitem

\bibitem[\protect\citeauthoryear{Skolnick et~al.}{2002}]{skolnick2002polariton}
\begin{barticle}
\bauthor{\bsnm{Skolnick}, \binits{M.}},
\bauthor{\bsnm{Stevenson}, \binits{R.}},
\bauthor{\bsnm{Tartakovskii}, \binits{A.}},
\bauthor{\bsnm{Butt{\'e}}, \binits{R.}},
\bauthor{\bsnm{Emam-Ismail}, \binits{M.}},
\bauthor{\bsnm{Whittaker}, \binits{D.}},
\bauthor{\bsnm{Savvidis}, \binits{P.}},
\bauthor{\bsnm{Baumberg}, \binits{J.}},
\bauthor{\bsnm{Lemaitre}, \binits{A.}},
\bauthor{\bsnm{Astratov}, \binits{V.}}, \betal:
\batitle{Polariton--polariton interactions and stimulated scattering in semiconductor microcavities}.
\bjtitle{Materials Science and Engineering: C}
\bvolume{19}(\bissue{1-2}),
\bfpage{407}--\blpage{416}
(\byear{2002})
\end{barticle}
\endbibitem

\bibitem[\protect\citeauthoryear{Su et~al.}{2021}]{su2021perovskite}
\begin{botherref}
\oauthor{\bsnm{Su}, \binits{R.}},
\oauthor{\bsnm{Fieramosca}, \binits{A.}},
\oauthor{\bsnm{Zhang}, \binits{Q.}},
\oauthor{\bsnm{Nguyen}, \binits{H.S.}},
\oauthor{\bsnm{Deleporte}, \binits{E.}},
\oauthor{\bsnm{Chen}, \binits{Z.}},
\oauthor{\bsnm{Sanvitto}, \binits{D.}},
\oauthor{\bsnm{Liew}, \binits{T.C.}},
\oauthor{\bsnm{Xiong}, \binits{Q.}}:
Perovskite semiconductors for room-temperature exciton-polaritonics.
Nature Materials,
1--10
(2021)
\end{botherref}
\endbibitem

\bibitem[\protect\citeauthoryear{Su et~al.}{2017}]{su2017room}
\begin{barticle}
\bauthor{\bsnm{Su}, \binits{R.}},
\bauthor{\bsnm{Diederichs}, \binits{C.}},
\bauthor{\bsnm{Wang}, \binits{J.}},
\bauthor{\bsnm{Liew}, \binits{T.C.}},
\bauthor{\bsnm{Zhao}, \binits{J.}},
\bauthor{\bsnm{Liu}, \binits{S.}},
\bauthor{\bsnm{Xu}, \binits{W.}},
\bauthor{\bsnm{Chen}, \binits{Z.}},
\bauthor{\bsnm{Xiong}, \binits{Q.}}:
\batitle{Room-temperature polariton lasing in all-inorganic perovskite nanoplatelets}.
\bjtitle{Nano letters}
\bvolume{17}(\bissue{6}),
\bfpage{3982}--\blpage{3988}
(\byear{2017})
\end{barticle}
\endbibitem

\bibitem[\protect\citeauthoryear{Su et~al.}{2018}]{su2018room}
\begin{barticle}
\bauthor{\bsnm{Su}, \binits{R.}},
\bauthor{\bsnm{Wang}, \binits{J.}},
\bauthor{\bsnm{Zhao}, \binits{J.}},
\bauthor{\bsnm{Xing}, \binits{J.}},
\bauthor{\bsnm{Zhao}, \binits{W.}},
\bauthor{\bsnm{Diederichs}, \binits{C.}},
\bauthor{\bsnm{Liew}, \binits{T.C.}},
\bauthor{\bsnm{Xiong}, \binits{Q.}}:
\batitle{Room temperature long-range coherent exciton polariton condensate flow in lead halide perovskites}.
\bjtitle{Science advances}
\bvolume{4}(\bissue{10}),
\bfpage{0244}
(\byear{2018})
\end{barticle}
\endbibitem

\bibitem[\protect\citeauthoryear{Fieramosca et~al.}{2019}]{fieramosca2019two}
\begin{barticle}
\bauthor{\bsnm{Fieramosca}, \binits{A.}},
\bauthor{\bsnm{Polimeno}, \binits{L.}},
\bauthor{\bsnm{Ardizzone}, \binits{V.}},
\bauthor{\bsnm{De~Marco}, \binits{L.}},
\bauthor{\bsnm{Pugliese}, \binits{M.}},
\bauthor{\bsnm{Maiorano}, \binits{V.}},
\bauthor{\bsnm{De~Giorgi}, \binits{M.}},
\bauthor{\bsnm{Dominici}, \binits{L.}},
\bauthor{\bsnm{Gigli}, \binits{G.}},
\bauthor{\bsnm{Gerace}, \binits{D.}}, \betal:
\batitle{Two-dimensional hybrid perovskites sustaining strong polariton interactions at room temperature}.
\bjtitle{Science advances}
\bvolume{5}(\bissue{5}),
\bfpage{9967}
(\byear{2019})
\end{barticle}
\endbibitem

\bibitem[\protect\citeauthoryear{Su et~al.}{2020}]{su2020observation}
\begin{barticle}
\bauthor{\bsnm{Su}, \binits{R.}},
\bauthor{\bsnm{Ghosh}, \binits{S.}},
\bauthor{\bsnm{Wang}, \binits{J.}},
\bauthor{\bsnm{Liu}, \binits{S.}},
\bauthor{\bsnm{Diederichs}, \binits{C.}},
\bauthor{\bsnm{Liew}, \binits{T.C.}},
\bauthor{\bsnm{Xiong}, \binits{Q.}}:
\batitle{Observation of exciton polariton condensation in a perovskite lattice at room temperature}.
\bjtitle{Nature Physics}
\bvolume{16}(\bissue{3}),
\bfpage{301}--\blpage{306}
(\byear{2020})
\end{barticle}
\endbibitem

\bibitem[\protect\citeauthoryear{Feng et~al.}{2021}]{feng2021all}
\begin{barticle}
\bauthor{\bsnm{Feng}, \binits{J.}},
\bauthor{\bsnm{Wang}, \binits{J.}},
\bauthor{\bsnm{Fieramosca}, \binits{A.}},
\bauthor{\bsnm{Bao}, \binits{R.}},
\bauthor{\bsnm{Zhao}, \binits{J.}},
\bauthor{\bsnm{Su}, \binits{R.}},
\bauthor{\bsnm{Peng}, \binits{Y.}},
\bauthor{\bsnm{Liew}, \binits{T.C.}},
\bauthor{\bsnm{Sanvitto}, \binits{D.}},
\bauthor{\bsnm{Xiong}, \binits{Q.}}:
\batitle{All-optical switching based on interacting exciton polaritons in self-assembled perovskite microwires}.
\bjtitle{Science Advances}
\bvolume{7}(\bissue{46}),
\bfpage{6627}
(\byear{2021})
\end{barticle}
\endbibitem

\bibitem[\protect\citeauthoryear{Belogur et~al.}{2022}]{belogur2022theory}
\begin{barticle}
\bauthor{\bsnm{Belogur}, \binits{A.}},
\bauthor{\bsnm{Baghdasaryan}, \binits{D.}},
\bauthor{\bsnm{Iorsh}, \binits{I.}},
\bauthor{\bsnm{Shelykh}, \binits{I.}},
\bauthor{\bsnm{Shahnazaryan}, \binits{V.}}:
\batitle{Theory of nonlinear excitonic response of hybrid organic perovskites in the regime of strong light-matter coupling}.
\bjtitle{Physical Review Applied}
\bvolume{17}(\bissue{4}),
\bfpage{044048}
(\byear{2022})
\end{barticle}
\endbibitem

\bibitem[\protect\citeauthoryear{Soufiani et~al.}{2015}]{soufiani2015polaronic}
\begin{botherref}
\oauthor{\bsnm{Soufiani}, \binits{A.M.}},
\oauthor{\bsnm{Huang}, \binits{F.}},
\oauthor{\bsnm{Reece}, \binits{P.}},
\oauthor{\bsnm{Sheng}, \binits{R.}},
\oauthor{\bsnm{Ho-Baillie}, \binits{A.}},
\oauthor{\bsnm{Green}, \binits{M.A.}}:
Polaronic exciton binding energy in iodide and bromide organic-inorganic lead halide perovskites.
Applied Physics Letters
\textbf{107}(23)
(2015)
\end{botherref}
\endbibitem

\bibitem[\protect\citeauthoryear{Baranowski and Plochocka}{2020}]{baranowski2020excitons}
\begin{barticle}
\bauthor{\bsnm{Baranowski}, \binits{M.}},
\bauthor{\bsnm{Plochocka}, \binits{P.}}:
\batitle{Excitons in metal-halide perovskites}.
\bjtitle{Advanced Energy Materials}
\bvolume{10}(\bissue{26}),
\bfpage{1903659}
(\byear{2020})
\end{barticle}
\endbibitem

\bibitem[\protect\citeauthoryear{Shi et~al.}{2020}]{shi2020exciton}
\begin{barticle}
\bauthor{\bsnm{Shi}, \binits{J.}},
\bauthor{\bsnm{Li}, \binits{Y.}},
\bauthor{\bsnm{Wu}, \binits{J.}},
\bauthor{\bsnm{Wu}, \binits{H.}},
\bauthor{\bsnm{Luo}, \binits{Y.}},
\bauthor{\bsnm{Li}, \binits{D.}},
\bauthor{\bsnm{Jasieniak}, \binits{J.J.}},
\bauthor{\bsnm{Meng}, \binits{Q.}}:
\batitle{Exciton character and high-performance stimulated emission of hybrid lead bromide perovskite polycrystalline film}.
\bjtitle{Advanced Optical Materials}
\bvolume{8}(\bissue{10}),
\bfpage{1902026}
(\byear{2020})
\end{barticle}
\endbibitem

\bibitem[\protect\citeauthoryear{Shang et~al.}{2018}]{shang2018surface}
\begin{barticle}
\bauthor{\bsnm{Shang}, \binits{Q.}},
\bauthor{\bsnm{Zhang}, \binits{S.}},
\bauthor{\bsnm{Liu}, \binits{Z.}},
\bauthor{\bsnm{Chen}, \binits{J.}},
\bauthor{\bsnm{Yang}, \binits{P.}},
\bauthor{\bsnm{Li}, \binits{C.}},
\bauthor{\bsnm{Li}, \binits{W.}},
\bauthor{\bsnm{Zhang}, \binits{Y.}},
\bauthor{\bsnm{Xiong}, \binits{Q.}},
\bauthor{\bsnm{Liu}, \binits{X.}}, \betal:
\batitle{Surface plasmon enhanced strong exciton--photon coupling in hybrid inorganic--organic perovskite nanowires}.
\bjtitle{Nano letters}
\bvolume{18}(\bissue{6}),
\bfpage{3335}--\blpage{3343}
(\byear{2018})
\end{barticle}
\endbibitem

\bibitem[\protect\citeauthoryear{Bouteyre et~al.}{2019}]{bouteyre2019room}
\begin{barticle}
\bauthor{\bsnm{Bouteyre}, \binits{P.}},
\bauthor{\bsnm{Nguyen}, \binits{H.S.}},
\bauthor{\bsnm{Lauret}, \binits{J.-S.}},
\bauthor{\bsnm{Tripp{\'e}-Allard}, \binits{G.}},
\bauthor{\bsnm{Delport}, \binits{G.}},
\bauthor{\bsnm{L{\'e}d{\'e}e}, \binits{F.}},
\bauthor{\bsnm{Diab}, \binits{H.}},
\bauthor{\bsnm{Belarouci}, \binits{A.}},
\bauthor{\bsnm{Seassal}, \binits{C.}},
\bauthor{\bsnm{Garrot}, \binits{D.}}, \betal:
\batitle{Room-temperature cavity polaritons with 3d hybrid perovskite: toward large-surface polaritonic devices}.
\bjtitle{ACS photonics}
\bvolume{6}(\bissue{7}),
\bfpage{1804}--\blpage{1811}
(\byear{2019})
\end{barticle}
\endbibitem

\bibitem[\protect\citeauthoryear{Masharin et~al.}{2023}]{masharin2023room}
\begin{barticle}
\bauthor{\bsnm{Masharin}, \binits{M.A.}},
\bauthor{\bsnm{Shahnazaryan}, \binits{V.A.}},
\bauthor{\bsnm{Iorsh}, \binits{I.V.}},
\bauthor{\bsnm{Makarov}, \binits{S.V.}},
\bauthor{\bsnm{Samusev}, \binits{A.K.}},
\bauthor{\bsnm{Shelykh}, \binits{I.A.}}:
\batitle{Room-temperature polaron-mediated polariton nonlinearity in mapbbr3 perovskites}.
\bjtitle{ACS Photonics}
\bvolume{10}(\bissue{3}),
\bfpage{691}--\blpage{698}
(\byear{2023})
\end{barticle}
\endbibitem

\bibitem[\protect\citeauthoryear{Masharin et~al.}{2022}]{masharin2022polaron}
\begin{barticle}
\bauthor{\bsnm{Masharin}, \binits{M.A.}},
\bauthor{\bsnm{Shahnazaryan}, \binits{V.A.}},
\bauthor{\bsnm{Benimetskiy}, \binits{F.A.}},
\bauthor{\bsnm{Krizhanovskii}, \binits{D.N.}},
\bauthor{\bsnm{Shelykh}, \binits{I.A.}},
\bauthor{\bsnm{Iorsh}, \binits{I.V.}},
\bauthor{\bsnm{Makarov}, \binits{S.V.}},
\bauthor{\bsnm{Samusev}, \binits{A.K.}}:
\batitle{Polaron-enhanced polariton nonlinearity in lead halide perovskites}.
\bjtitle{Nano Letters}
\bvolume{22},
\bfpage{9092}--\blpage{9099}
(\byear{2022})
\end{barticle}
\endbibitem

\bibitem[\protect\citeauthoryear{Masharin et~al.}{2023}]{masharin2023lasing}
\begin{barticle}
\bauthor{\bsnm{Masharin}, \binits{M.A.}},
\bauthor{\bsnm{Samusev}, \binits{A.}},
\bauthor{\bsnm{Bogdanov}, \binits{A.}},
\bauthor{\bsnm{Iorsh}, \binits{I.}},
\bauthor{\bsnm{Demir}, \binits{H.V.}},
\bauthor{\bsnm{Makarov}, \binits{S.}}:
\batitle{Room-temperature exceptional-point-driven polariton lasing from perovskite metasurface}.
\bjtitle{Advanced Functional Materials}
\bvolume{33}(\bissue{22}),
\bfpage{2215007}
(\byear{2023})
\end{barticle}
\endbibitem

\bibitem[\protect\citeauthoryear{Masharin et~al.}{2024}]{masharin2024giant}
\begin{barticle}
\bauthor{\bsnm{Masharin}, \binits{M.A.}},
\bauthor{\bsnm{Oskolkova}, \binits{T.}},
\bauthor{\bsnm{Isik}, \binits{F.}},
\bauthor{\bsnm{Volkan~Demir}, \binits{H.}},
\bauthor{\bsnm{Samusev}, \binits{A.K.}},
\bauthor{\bsnm{Makarov}, \binits{S.V.}}:
\batitle{Giant ultrafast all-optical modulation based on exceptional points in exciton--polariton perovskite metasurfaces}.
\bjtitle{ACS nano}
\bvolume{18}(\bissue{4}),
\bfpage{3447}--\blpage{3455}
(\byear{2024})
\end{barticle}
\endbibitem

\bibitem[\protect\citeauthoryear{Laitz et~al.}{2023}]{laitz2023uncovering}
\begin{barticle}
\bauthor{\bsnm{Laitz}, \binits{M.}},
\bauthor{\bsnm{Kaplan}, \binits{A.E.}},
\bauthor{\bsnm{Deschamps}, \binits{J.}},
\bauthor{\bsnm{Barotov}, \binits{U.}},
\bauthor{\bsnm{Proppe}, \binits{A.H.}},
\bauthor{\bsnm{Garc{\'\i}a-Benito}, \binits{I.}},
\bauthor{\bsnm{Osherov}, \binits{A.}},
\bauthor{\bsnm{Grancini}, \binits{G.}},
\bauthor{\bsnm{deQuilettes}, \binits{D.W.}},
\bauthor{\bsnm{Nelson}, \binits{K.A.}}, \betal:
\batitle{Uncovering temperature-dependent exciton-polariton relaxation mechanisms in hybrid organic-inorganic perovskites}.
\bjtitle{Nature Communications}
\bvolume{14}(\bissue{1}),
\bfpage{2426}
(\byear{2023})
\end{barticle}
\endbibitem

\bibitem[\protect\citeauthoryear{Wu et~al.}{2021}]{wu2021nonlinear}
\begin{barticle}
\bauthor{\bsnm{Wu}, \binits{J.}},
\bauthor{\bsnm{Ghosh}, \binits{S.}},
\bauthor{\bsnm{Su}, \binits{R.}},
\bauthor{\bsnm{Fieramosca}, \binits{A.}},
\bauthor{\bsnm{Liew}, \binits{T.C.}},
\bauthor{\bsnm{Xiong}, \binits{Q.}}:
\batitle{Nonlinear parametric scattering of exciton polaritons in perovskite microcavities}.
\bjtitle{Nano Letters}
\bvolume{21}(\bissue{7}),
\bfpage{3120}--\blpage{3126}
(\byear{2021})
\end{barticle}
\endbibitem

\bibitem[\protect\citeauthoryear{Masharin et~al.}{2024}]{masharin2024polariton}
\begin{barticle}
\bauthor{\bsnm{Masharin}, \binits{M.A.}},
\bauthor{\bsnm{Khmelevskaia}, \binits{D.}},
\bauthor{\bsnm{Kondratiev}, \binits{V.I.}},
\bauthor{\bsnm{Markina}, \binits{D.I.}},
\bauthor{\bsnm{Utyushev}, \binits{A.D.}},
\bauthor{\bsnm{Dolgintsev}, \binits{D.M.}},
\bauthor{\bsnm{Dmitriev}, \binits{A.D.}},
\bauthor{\bsnm{Shahnazaryan}, \binits{V.A.}},
\bauthor{\bsnm{Pushkarev}, \binits{A.P.}},
\bauthor{\bsnm{Isik}, \binits{F.}}, \betal:
\batitle{Polariton lasing in mie-resonant perovskite nanocavity}.
\bjtitle{Opto-Electronic Advances}
\bvolume{7}(\bissue{4}),
\bfpage{230148}
(\byear{2024})
\end{barticle}
\endbibitem

\bibitem[\protect\citeauthoryear{Alias et~al.}{2016}]{alias2016optical}
\begin{barticle}
\bauthor{\bsnm{Alias}, \binits{M.S.}},
\bauthor{\bsnm{Dursun}, \binits{I.}},
\bauthor{\bsnm{Saidaminov}, \binits{M.I.}},
\bauthor{\bsnm{Diallo}, \binits{E.M.}},
\bauthor{\bsnm{Mishra}, \binits{P.}},
\bauthor{\bsnm{Ng}, \binits{T.K.}},
\bauthor{\bsnm{Bakr}, \binits{O.M.}},
\bauthor{\bsnm{Ooi}, \binits{B.S.}}:
\batitle{Optical constants of ch 3 nh 3 pbbr 3 perovskite thin films measured by spectroscopic ellipsometry}.
\bjtitle{Optics express}
\bvolume{24}(\bissue{15}),
\bfpage{16586}--\blpage{16594}
(\byear{2016})
\end{barticle}
\endbibitem

\bibitem[\protect\citeauthoryear{Kondratyev et~al.}{2023}]{kondratyev2023probing}
\begin{barticle}
\bauthor{\bsnm{Kondratyev}, \binits{V.I.}},
\bauthor{\bsnm{Permyakov}, \binits{D.V.}},
\bauthor{\bsnm{Ivanova}, \binits{T.V.}},
\bauthor{\bsnm{Iorsh}, \binits{I.V.}},
\bauthor{\bsnm{Krizhanovskii}, \binits{D.N.}},
\bauthor{\bsnm{Skolnick}, \binits{M.S.}},
\bauthor{\bsnm{Kravtsov}, \binits{V.}},
\bauthor{\bsnm{Samusev}, \binits{A.K.}}:
\batitle{Probing and control of guided exciton--polaritons in a 2d semiconductor-integrated slab waveguide}.
\bjtitle{Nano Letters}
\bvolume{23}(\bissue{17}),
\bfpage{7876}--\blpage{7882}
(\byear{2023})
\end{barticle}
\endbibitem

\bibitem[\protect\citeauthoryear{Guo et~al.}{2017}]{guo2017polar}
\begin{barticle}
\bauthor{\bsnm{Guo}, \binits{P.}},
\bauthor{\bsnm{Xia}, \binits{Y.}},
\bauthor{\bsnm{Gong}, \binits{J.}},
\bauthor{\bsnm{Stoumpos}, \binits{C.C.}},
\bauthor{\bsnm{McCall}, \binits{K.M.}},
\bauthor{\bsnm{Alexander}, \binits{G.C.}},
\bauthor{\bsnm{Ma}, \binits{Z.}},
\bauthor{\bsnm{Zhou}, \binits{H.}},
\bauthor{\bsnm{Gosztola}, \binits{D.J.}},
\bauthor{\bsnm{Ketterson}, \binits{J.B.}}, \betal:
\batitle{Polar fluctuations in metal halide perovskites uncovered by acoustic phonon anomalies}.
\bjtitle{ACS Energy Letters}
\bvolume{2}(\bissue{10}),
\bfpage{2463}--\blpage{2469}
(\byear{2017})
\end{barticle}
\endbibitem

\bibitem[\protect\citeauthoryear{Butler et~al.}{2015}]{butler2015band}
\begin{barticle}
\bauthor{\bsnm{Butler}, \binits{K.T.}},
\bauthor{\bsnm{Frost}, \binits{J.M.}},
\bauthor{\bsnm{Walsh}, \binits{A.}}:
\batitle{Band alignment of the hybrid halide perovskites ch3nh3pbcl3, ch3nh3pbbr3 and ch3nh3pbi3}.
\bjtitle{Materials Horizons}
\bvolume{2}(\bissue{2}),
\bfpage{228}--\blpage{231}
(\byear{2015})
\end{barticle}
\endbibitem

\bibitem[\protect\citeauthoryear{Mante et~al.}{2017}]{mante2017electron}
\begin{barticle}
\bauthor{\bsnm{Mante}, \binits{P.-A.}},
\bauthor{\bsnm{Stoumpos}, \binits{C.C.}},
\bauthor{\bsnm{Kanatzidis}, \binits{M.G.}},
\bauthor{\bsnm{Yartsev}, \binits{A.}}:
\batitle{Electron--acoustic phonon coupling in single crystal ch3nh3pbi3 perovskites revealed by coherent acoustic phonons}.
\bjtitle{Nature communications}
\bvolume{8}(\bissue{1}),
\bfpage{14398}
(\byear{2017})
\end{barticle}
\endbibitem

\bibitem[\protect\citeauthoryear{Jeon et~al.}{2014}]{jeon2014solvent}
\begin{barticle}
\bauthor{\bsnm{Jeon}, \binits{N.J.}},
\bauthor{\bsnm{Noh}, \binits{J.H.}},
\bauthor{\bsnm{Kim}, \binits{Y.C.}},
\bauthor{\bsnm{Yang}, \binits{W.S.}},
\bauthor{\bsnm{Ryu}, \binits{S.}},
\bauthor{\bsnm{Seok}, \binits{S.I.}}:
\batitle{Solvent engineering for high-performance inorganic--organic hybrid perovskite solar cells}.
\bjtitle{Nature materials}
\bvolume{13}(\bissue{9}),
\bfpage{897}--\blpage{903}
(\byear{2014})
\end{barticle}
\endbibitem

\end{thebibliography}
\end{document}